\input{aipcheck}

\documentclass[
    ,final            
  ]
  {aipproc}

\layoutstyle{6x9}

\begin{document}

%
%
%

\def\aj{AJ}%
\def\araa{ARA\&A}%
\def\apj{ApJ}%
\def\apjl{ApJ}%
\def\apjs{ApJS}%
\def\ao{Appl.~Opt.}%
\def\apss{Ap\&SS}%
\def\aap{A\&A}%
\def\aapr{A\&A~Rev.}%
\def\aaps{A\&AS}%
\def\azh{AZh}%
\def\baas{BAAS}%
\def\jrasc{JRASC}%
\def\memras{MmRAS}%
\def\mnras{MNRAS}%
\def\pra{Phys.~Rev.~A}%
\def\prb{Phys.~Rev.~B}%
\def\prc{Phys.~Rev.~C}%
\def\prd{Phys.~Rev.~D}%
\def\pre{Phys.~Rev.~E}%
\def\prl{Phys.~Rev.~Lett.}%
\def\pasp{PASP}%
\def\pasj{PASJ}%
\def\qjras{QJRAS}%
\def\skytel{S\&T}%
\def\solphys{Sol.~Phys.}%
\def\sovast{Soviet~Ast.}%
\def\ssr{Space~Sci.~Rev.}%
\def\zap{ZAp}%
\def\nat{Nature}%
\def\iaucirc{IAU~Circ.}%
\def\aplett{Astrophys.~Lett.}%
\def\apspr{Astrophys.~Space~Phys.~Res.}%
\def\bain{Bull.~Astron.~Inst.~Netherlands}%
\def\fcp{Fund.~Cosmic~Phys.}%
\def\gca{Geochim.~Cosmochim.~Acta}%
\def\grl{Geophys.~Res.~Lett.}%
\def\jcp{J.~Chem.~Phys.}%
\def\jgr{J.~Geophys.~Res.}%
\def\jqsrt{J.~Quant.~Spec.~Radiat.~Transf.}%
\def\memsai{Mem.~Soc.~Astron.~Italiana}%
\def\nphysa{Nucl.~Phys.~A}%
\def\physrep{Phys.~Rep.}%
\def\physscr{Phys.~Scr}%
\def\planss{Planet.~Space~Sci.}%
\def\procspie{Proc.~SPIE}%
\let\astap=\aap
\let\apjlett=\apjl
\let\apjsupp=\apjs
\let\applopt=\ao

\newcommand{\xmm}{XMM-{\em Newton}}
\newcommand{\lesssim}{<\sim}   
\newcommand{\gtrsim}{>\sim}
\newcommand{\de}{{\rm d}}
\newcommand{\e}[1]{\cdot 10^{#1}}

\title{X-ray gaseous emission in the galaxy M82}


\author{Piero Ranalli}{
  address={Università di Bologna, via Ranzani 1, 40127 Bologna, Italy}
}

\author{Andrea Comastri}{
  address={INAF -- Osservatorio Astronomico di Bologna, via Ranzani 1, 40127 Bologna, Italy}
}

\author{Livia Origlia}{
  address={INAF -- Osservatorio Astronomico di Bologna, via Ranzani 1, 40127 Bologna, Italy}
}

\author{Roberto Maiolino}{
  address={INAF -- Osservatorio Astronomico di Roma, via di Frascati 33, 00040 Roma, Italy}
}

\begin{abstract}
 The main results from a deep X-ray observation of M82 are
  summarised: spatially-dependent chemical abundances, temperature
  structure of the gas, charge-exchange emission lines in the
  spectrum. We also present an update of the chemical abundances,
  based on a more refined extraction of spectra.
\end{abstract}

\maketitle


\section{Introduction}

\begin{table}
  \begin{tabular}{lrrrrrrrrrrr}
\hline  
 &\tablehead{1}{r}{b}{N5}
 &\tablehead{1}{r}{b}{N4}
 &\tablehead{1}{r}{b}{N3}
 &\tablehead{1}{r}{b}{N2}
 &\tablehead{1}{r}{b}{N1}
 &\tablehead{1}{r}{b}{centre}
 &\tablehead{1}{r}{b}{S1}
 &\tablehead{1}{r}{b}{S2}
 &\tablehead{1}{r}{b}{S3}
 &\tablehead{1}{r}{b}{S4}
 &\tablehead{1}{r}{b}{S5}\\
\hline
  Normalisation ($10^{-5}$)
                       &0.079  &0.24  &0.50 &0.41 &0.85  &27     &1.7  &0.92  &0.20  &0.099 &0.047 \\
  Volume (kpc$^3$)     &0.61   &1.2   &1.1  &0.43 &0.41  &1.4    &0.24 &0.35  &0.26  &0.21  &0.22
                                                                                         \smallskip \\
  Density ($10^{-3}$ cm$^{-3}$)
                       &$2.6 f^{-\frac{1}{2}}$
                               &3.3   &4.8  &7.1  &10    &32     &20   &12    &6.4   &5.1   &3.3  \\
  Pressure ($10^{-12}$ dine cm$^{-2}$)
                       &$2.8 f^{-\frac{1}{2}}$
                               &3.5   &5.7  &10   &18    &54     &32   &19    &9.2   &6.1   &3.6  \\
  Mass ($10^5$ M$_\odot$)
                       &$0.40  f^{+\frac{1}{2}}$
                               &0.99  &1.3  &0.76 &1.1   &11     &1.1  &1.0   &0.42  &0.26  &0.19 \\
  Energy ($10^{53}$ erg)
                       &$0.75  f^{+\frac{1}{2}}$
                               &1.9   &2.9  &1.9  &3.2   &33     &3.3  &2.9   &1.1   &0.56  &0.36 \\
  Energy density ($10^{-12}$ erg cm$^{-3}$)
                       &$4.1   f^{+\frac{1}{2}}$
                               &5.3   &8.6  &15   &27    &81     &48   &28    &14    &9.2   &5.4   \\
  Cooling time (Myr)
                       &$600   f^{+\frac{1}{2}}$
                               &500   &380  &360  &300   &100    &150  &250   &400   &380   &490  \\
\hline
  \end{tabular}
  \caption{Physical parameters of the plasma across the different
    regions of the outflow. The dependencies on the filling factor $f$
    have been explicited for simplicity only in the first column, but
    they apply to all columns.}
  \label{tab:fisica}
\end{table}

We performed a very deep (100 ks) observation of the starburst galaxy
M82 with the EPIC and RGS instruments on-board \xmm.  The analysis has
been published in \cite{m82centok}; we refer for any detail to that
paper.  A brief summary of the main results is presented in the
following.

M82 has a large and luminous outflow, extending for several kpc out of
both sides of the galaxy plane. On the sky, it extends for about 10
arcmin.  A good-quality spectrum was obtained with the RGS
instrument. However, the main difficulty in analysing X-ray grating
spectra from extended sources is the dependence of the
line-spread-function from the spatial shape of the source. The rgsxsrc
model which convolves the spectral response with the shape of the
source 
 was essential to this
analysis. Also, the parts of the outflow which lie inside of the
galactic disc are heavily absorbed: thus the outflow appears with a
different shape at different energies. For this reason, we had to
analyse separately the long- (18--30 \AA) and short-wavelength (6-18
\AA) regions of the spectrum.

The EPIC instrument allowed the spatially-resolved spectroscopy of
the outflow, albeit at a lower resolution than RGS.
The outflow was divided in eleven slices, each one 
parallel to the galactic plane, and spectra were extracted and
analysed for each slice, deriving the temperature structure,
abundances, and physical parameters of the plasma.

Finally, we publish here update figures and table: we refined the
extraction regions, after discovering that the regions used for spectral extractions in
\cite{m82centok} were actually larger than intended. 
No significant change in any conclusion or derived value has
occurred. 

\section{Main results}

At least three spectral components are present
in the broad-band spectrum: i) continuum emission from point sources;
ii) thermal plasma emission from hot gas; iii) charge exchange
emission from neutral metals (Mg and Si). 

The chemical absolute abundances of the thermal plasma depend on the
distance from the galactic plane: they are larger in the outskirts and
smaller close to the galaxy centre. The abundance ratios also show
spatial variations (see below). This might be due to the
dependence of supernova (SN) yields on the progenitor mass, if the
matter expelled by the first SN in a star formation burst  is also
the first one to travel out of the galactic plane. It may also
be a clear example of metals being pushed in the inter-galactic
medium.

The X-ray derived
Oxygen abundance is lower than that measured in the atmospheres of red
supergiant stars, leading to the hypothesis that a significant
fraction of Oxygen ions has already cooled off and no longer emits at
energies $\gtrsim 0.5$ keV.

Two lines were found in the EPIC and RGS spectra, significant at a confidence
level $>99.97\%$, which may be attributed to charge-exchange (CE)
emission\cite{chamberlain56} . A third one was found in the EPIC spectra only. 
The RGS spectrum of the O VII He-like triplet is consistent with CE
emission, but not with a thermal origin. The CE process is usually
observed in cometary and planetary emission, and it is due to
the interaction of an energetic ion with a dust
grain. It might also be one of the cooling channels for O ions.

The differential emission measure (DEM) of the thermal plasma was
found to have a bimodal structure, with one peak around 0.5 keV and
slightly dependent on the distance from the galactic plane, and
another peak around 6 keV. The currently favoured explanation for the
double-peaked structure is the Masai model for plasma emission
\cite{masaidogiel02,dogielmasai02}: if the plasma emission occurs in
the same region where particle are accelerated, then the electron
energy distribution is not fully thermal, and the spectrum emitted by
the non-thermal part has a defined spectral shape which can be fitted as a
higher-temperature plasma. Another possible explanation ---that the
$\sim 6$ keV plasma is actually non-thermal emission from unresolved
point sources--- is difficult to explain, because it would require a
sizable population of point sources lying many kpc above the galactic
plane.

\section{Spatially-dependent chemical abundances}

We defined 11 regions in the outflow of M82, slicing it parallel
to the galactic plane.  In the
figures, negative and positive heights above the galactic plane are
assigned to the northern and southern regions, respectively.  The
spectral model was a multi-temperature APEC plasma, with temperatures
ranging from 0.1 to 10 keV. Point sources were excluded from all
spectral regions except the galaxy centre, where their contribution was
added to the model in the form of an absorbed power-law.  The
best-fitting abundances are shown in Fig.~1 (left panels) along with
results from infrared observations for the central regions \cite{or04}
which show the abundances of stars born before the start of the
current burst. Lighter $\alpha$-elements are more concentrated in the
outflow than in the centre. This effect is larger for elements with
lower atomic mass, becomes less evident for Si and reverses for S.
The centre/outskirt abundance ratio in the centre is about $\sim 1/10$
for O and Ne. Fe is also more concentrated in the outflow.

The abundances ratios (Fig.~1 right panel) have smaller
variations, and present different trends for light and massive
elements: while the O/Fe and Ne/Fe ratios are lower in the centre than
in the outskirts, the opposite holds for Si/Fe and S/Fe, with Mg/Fe
being an intermediate case showing no variation. The scatter between
values for centre and outskirts is a factor $\lesssim 3$.

\begin{figure*}[h]
\centering
\includegraphics[width=9.7cm,bb=165 300 444 716,clip]{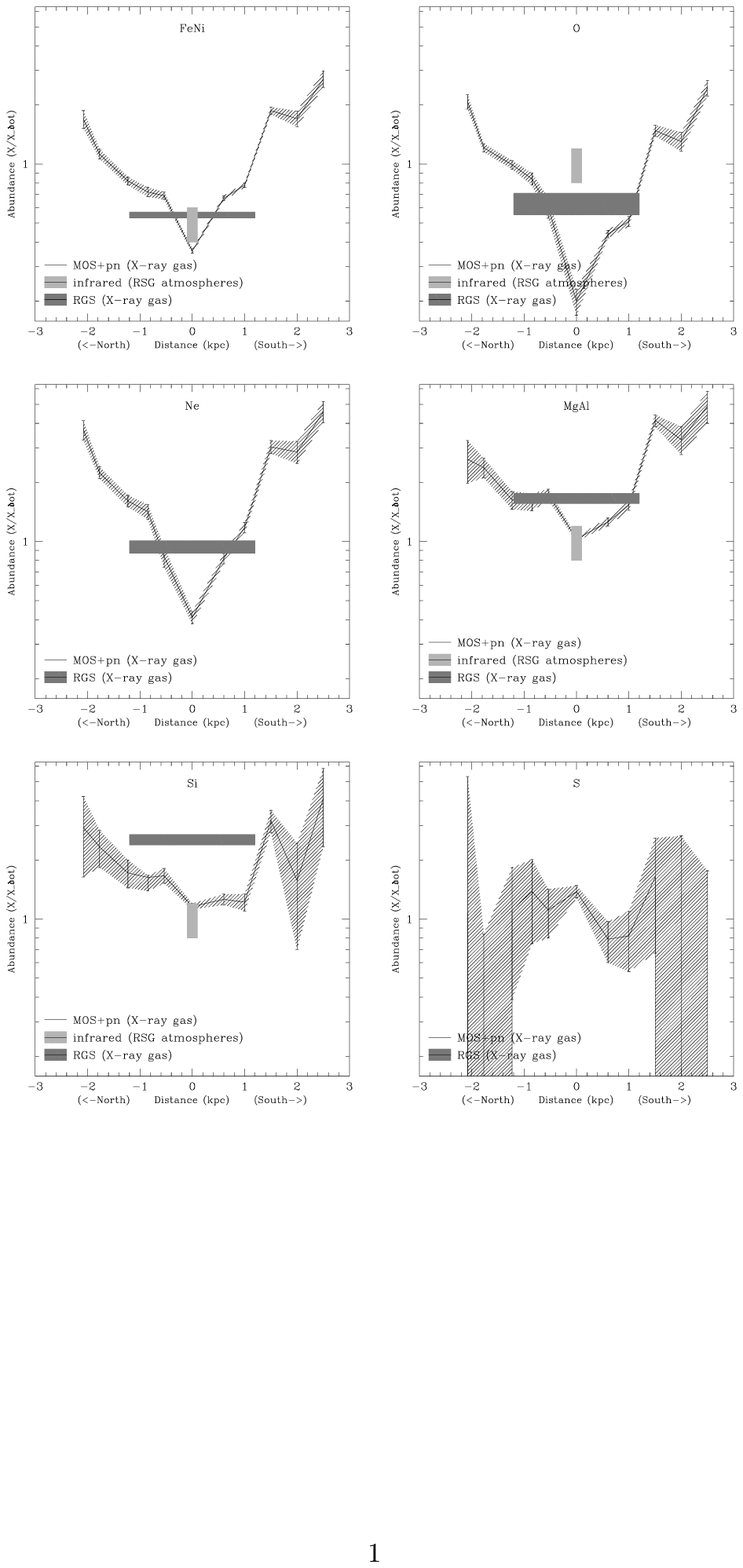}
\includegraphics[width=6.7cm]{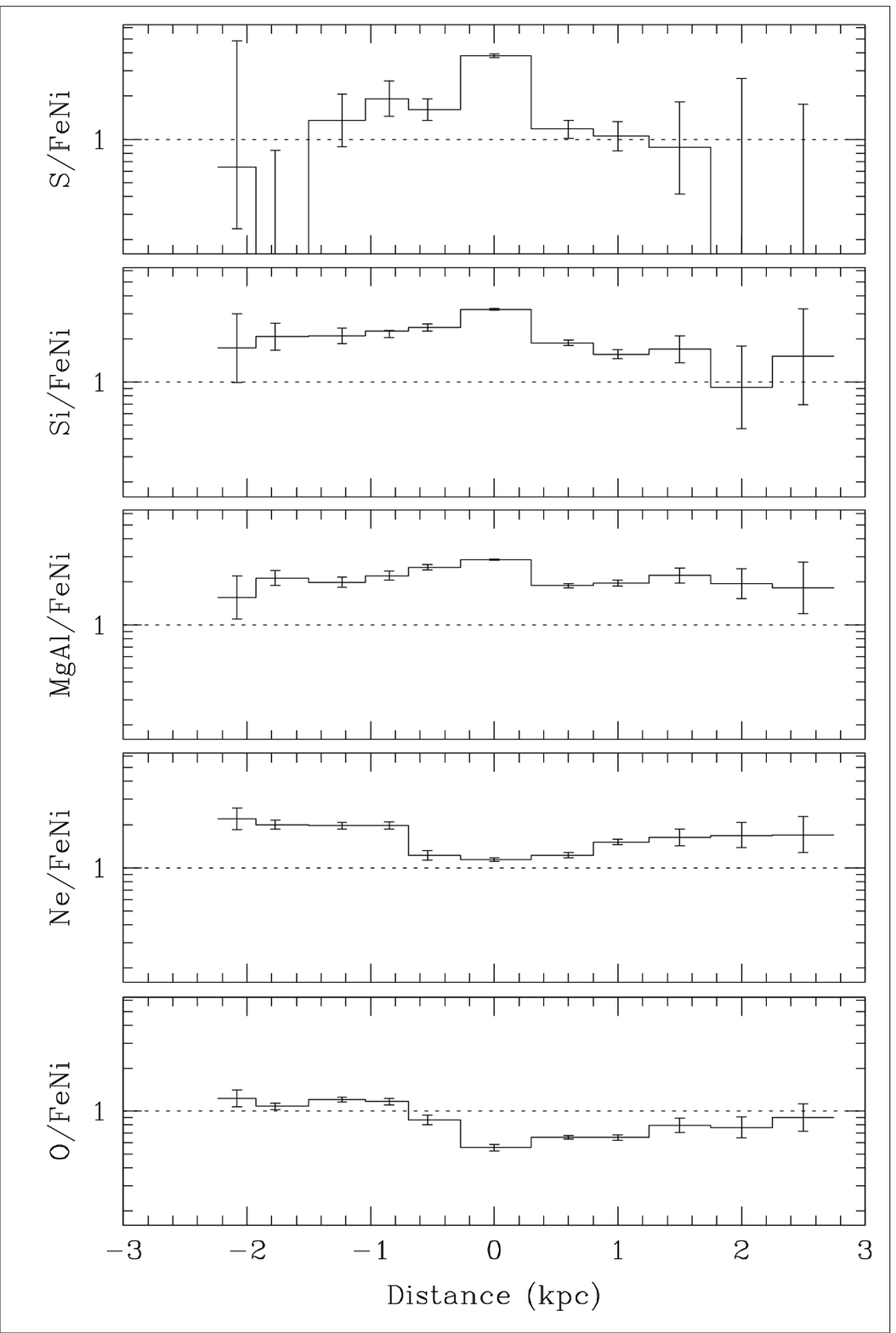}
\caption{ {\em Left panels:} Variation of chemical abundances with increasing height on
    the galactic plane. Black: abundances from X-ray MOS and {\em pn}
    data. Dark grey: abundances from X-ray RGS data (due to the
    characteristics of the RGS spectrometer, they represent
    space-averaged values). Light grey: abundances from infrared data
    (corresponding to red supergiant stars in the galaxy central
    region). Negative values of distance refer to the north direction,
    positive values to south.\newline
{\em Right panel:} 
Abundance ratios (X/Fe) observed in EPIC spectra of the
    outflow. Negative values of height refer to the north
    direction, positive values to south.
 }
\end{figure*}

The physical parameters of the plasma may be obtained from the
best-fitting temperature and normalisation of the model, with some
assumptions about the volume and filling factor.
From Table~1, one may see that the gas density and
pressure decrease by a factor of $\sim 10$ from the centre to the
outskirts, while the cooling time increases.





\bibliographystyle{aipproc}   

\bibliography{../fullbiblio}

\IfFileExists{\jobname.bbl}{}
 {\typeout{}
  \typeout{******************************************}
  \typeout{** Please run "bibtex \jobname" to optain}
  \typeout{** the bibliography and then re-run LaTeX}
  \typeout{** twice to fix the references!}
  \typeout{******************************************}
  \typeout{}
 }

\end{document}